\documentclass[preprint,11pt]{elsarticle}
\biboptions{sort&compress}
\usepackage{graphicx}
\usepackage{amscd}
\usepackage{amsmath}
\usepackage{amssymb}
\usepackage{amsfonts}
\usepackage{verbatim}
\usepackage{mathptmx}
\usepackage{setspace}
\usepackage{url}
\usepackage{multirow}
\usepackage{algorithm}
\usepackage{algpseudocode}
\usepackage{float}
\usepackage[normalem]{ulem}
\usepackage{caption}
\usepackage{subcaption}
\usepackage{color}
\usepackage{siunitx}
\usepackage[a4paper,
left=2.0cm,
right=2.0cm,
top=2.5cm,
bottom=2.5cm]{geometry}

\pdfoutput=1

\begin{document}

\journal{Physica A}

\begin{frontmatter}

\title{Patterns in a Warming Ocean: Stylized Spectral Facts on Sea Surface Temperature}

\author[ufrgs]{Roberto da Silva\corref{cor1}}
\ead{rdasilva@if.ufrgs.br}

\author[ufrgs]{Eliseu Venites Filho}
\ead{eliseuv816@gmail.com}

\author[ufrgs]{Eduardo V. Stock}
\ead{eduardo.stock@ufrgs.br}

\author[ufrgs]{Sebastian Gonçalves}
\ead{sconc@if.ufrgs.br}

\cortext[cor1]{Corresponding author}

\address[ufrgs]{
Instituto de F\'{\i}sica, Universidade Federal do Rio Grande do Sul (UFRGS),\\
Caixa Postal 15.051 -- Porto Alegre -- RS -- Brazil
}

\begin{abstract}
Capturing stylized facts and their evolution is essential for understanding the impact of climate change on complex environmental systems. In this work, we investigate the spectral properties of correlation matrices constructed from sea surface temperature data, employing tools from random matrix theory to identify signatures of global warming and long-term climate variability. By constructing yearly ensembles of correlation matrices, we analyze the evolution of both the eigenvalue density and the statistical behavior of the largest eigenvalue. Our results reveal significant departures from the universal behavior expected for random correlations. In particular, the empirical spectra systematically deviate from the Marchenko--Pastur law, indicating the presence of strong collective correlations in ocean temperature dynamics. Moreover, we find that the average largest eigenvalue exhibits a pronounced increase over time, closely following the rise in global mean ocean temperature. The distribution of the largest eigenvalue is found to be approximately Gaussian rather than Tracy--Widom, suggesting that the system lies outside the standard universal regime associated with weakly correlated Wishart ensembles. Together with conventional statistical indicators, these spectral signatures consistently reflect the progressive modification of the ocean-temperature correlation structure. Our findings demonstrate that spectral observables derived from correlation matrices, particularly the largest eigenvalue and its fluctuations, provide a sensitive framework for characterizing climate dynamics and detecting emerging signatures of long-term climate change.
\end{abstract}

\end{frontmatter}

\section{Introduction}

The existence of anthropogenic global warming is supported by a broad body of theoretical, observational, and attribution studies. Since the pioneering work of Arrhenius, who first quantified the influence of atmospheric carbon dioxide on Earth's temperature \cite{arrhenius1896influence}, successive investigations have progressively strengthened the evidence linking increasing greenhouse gas concentrations to long-term climate change. Observational records, including the measurements initiated by Keeling \cite{keeling1960concentration}, together with attribution studies based on climate observations and numerical models, have established that human influence has become the dominant driver of the
observed warming trend \cite{Santer1996,Hansen2010,IPCC_2021}.

Among the various indicators of climate change, sea surface temperature (SST) has received particular attention because the oceans play a central role in the Earth's energy balance and heat transport\cite{EarthsGlobalEnergyBudget,Science2017_Cheng,Cheng2024}. Beyond their thermodynamic importance, oceanic temperature fields contain signatures of large-scale collective processes associated with climate variability, including basin-scale oscillations\cite{Mantua_AMS_1997,Enfield_GRL_2001}, teleconnection patterns\cite{Wallace_AMS_1981,Trenberth_1998}, and long-term warming trends \cite{Trenberth_AMS_1997,Cheng2020,2025_Ghadamidehno}. As a consequence, SST records constitute a paradigmatic example of a spatiotemporal geophysical field in which observations collected at different locations are intrinsically correlated across multiple spatial and temporal scales.

The analysis of covariance structures arising from such fields has a long tradition in climate science\cite{preisendorfer1988principal,Storch_Zwiers_1999,wilks2006statistical}. Since the pioneering work of Lorenz \cite{lorenz1956empirical}, Empirical Orthogonal Functions (EOFs) and their equivalent formulation through Principal Component Analysis (PCA) have become standard tools for identifying dominant modes of climate variability. These techniques decompose covariance matrices into orthogonal modes and have been widely employed in the study of oceanic and atmospheric phenomena. However, while EOF analysis provides a powerful framework for dimensionality reduction, the interpretation of its modes requires caution. Sampling effects, mode degeneracies, and the mathematical constraints imposed by orthogonality may complicate the association between empirical modes and physically independent climatic processes \cite{EOF_North1982,EOF_Hannachi_2007,EOF_Adam2009}.

More recently, complementary perspectives based on complex networks and spectral methods have been developed to investigate the statistical organization of climate data. In particular, the relationship between eigenvalue-based approaches and climate network methodologies has been explored as a means of characterizing large-scale climate variability from different but closely related viewpoints \cite{TSONIS2004497,Donges2015}. These developments suggest that the eigenvalue spectrum of climatic covariance matrices may contain valuable information beyond the leading EOF modes traditionally examined in climatological studies.

From the perspective of Random Matrix Theory (RMT), covariance matrices naturally arise within the framework of Wishart ensembles \cite{Wishart1928}. In the limit of uncorrelated variables and large matrix dimensions, the eigenvalue density is described by the Marchenko--Pastur (MP) law \cite{marchenko1967distribution}, which provides a benchmark for distinguishing statistical noise from genuine correlation structures. Deviations from this reference behavior have attracted considerable interest in fields ranging from condensed matter physics to quantitative finance \cite{PhysRevLett.83.1467,PhysRevLett.83.1471,stanleyQuantifyingFluctuationsEconomic2000,bouchaudFinancialRMT2009}, where outlier eigenvalues are often associated with collective modes that cannot be explained solely by finite-sampling fluctuations.

The present work is particularly motivated by recent applications of correlation-matrix methods in statistical and nonlinear physics. Previous studies have shown that matrices constructed from ensembles of stochastic realizations can be used to identify critical points, chaotic regimes, and emergent collective phenomena in systems with and without well-defined Hamiltonians \cite{daSilvaRandomMatricesTheory2023,daSilva2023meanField,nosso_article_entropy_2024,daSilva2024PottsModel,daSilva2025Identifying,daSilva2025RevisitingContact,daSilva2023Dynamics}. In those investigations, the matrix entries were obtained from independent realizations generated under identical model parameters, allowing the spectral properties of the correlation matrix to reveal changes in the collective behavior of the underlying system.

The climatic problem considered here differs in an important aspect. Rather than representing repeated realizations of the same stochastic process, the time series correspond to measurements collected at different spatial locations of a geophysical field. Consequently, the resulting covariance matrices encode not only statistical fluctuations but also the intrinsic spatial and temporal correlations of the climate system. In this sense, the present setting is more closely related to correlated Wishart ensembles \cite{Burda2005}, where the underlying variables may exhibit nontrivial correlation structures across both space and time. Such models provide a useful conceptual framework for understanding how empirical covariance spectra are influenced by genuine correlations as well as by finite-sampling effects.

The statistical properties of extreme eigenvalues have also received considerable attention within RMT. In the null hypothesis of purely random covariance matrices, the fluctuations of the largest eigenvalue are governed by Tracy--Widom statistics \cite{Johnstone2000}. More generally, studies of spiked covariance models have shown that sufficiently strong collective modes may cause dominant eigenvalues to detach from the bulk spectrum, producing outlier modes associated with underlying correlation structures \cite{BBP2005}. Although the present work does not attempt to establish a direct correspondence with these theoretical models, such results provide an important conceptual framework for interpreting the emergence of dominant spectral modes in empirical climate data.

In this paper, we investigate whether covariance matrices constructed from SST time series exhibit robust spectral features associated with climate variability and long-term warming. Rather than focusing exclusively on the dominant modes identified by traditional EOF analyses, we examine the statistical properties of the entire eigenvalue spectrum within an RMT framework. Our central objective is to determine whether the evolution of the spectral structure contains signatures of the large-scale reorganization of oceanic temperature correlations over recent decades, thereby providing complementary information about the collective behavior of the climate system.

The remainder of this paper is organized as follows. First, we present the dataset and describe the preprocessing procedure adopted to remove seasonal variability from the SST records. Next, we introduce the construction of the correlation matrices and the main concepts of the spectral analysis in Section~\ref{Sec:random_matrices_methodology}. The spectral properties of the resulting matrices, including deviations from the MP distribution and the evolution of the largest eigenvalues, are discussed in Section~\ref{Sec:Spectral_Analysis}. Finally, our conclusions and perspectives are summarized in Section~\ref{Sec:Conclusions}.

\section{The Dataset and Descriptive Statistics}

\label{Sec:dataset}

The Daily Optimum Interpolation Sea Surface Temperature (DOISST) v2.1
dataset \cite{NOAA21}, provided by the National Oceanic and Atmospheric
Administration (NOAA), combines in situ (ships and buoys) and satellite
observations. The creation process addresses biases inherent to each method
and uses interpolation to produce a consistent daily record. This results in
sea surface temperature data on a geospatial mesh with a $\SI{0.25}{\degree}$
resolution for both latitude and longitude, spanning from September 1st,
1981 to the present.

\begin{figure}[tbp]
\centering\includegraphics[width=0.9\linewidth]{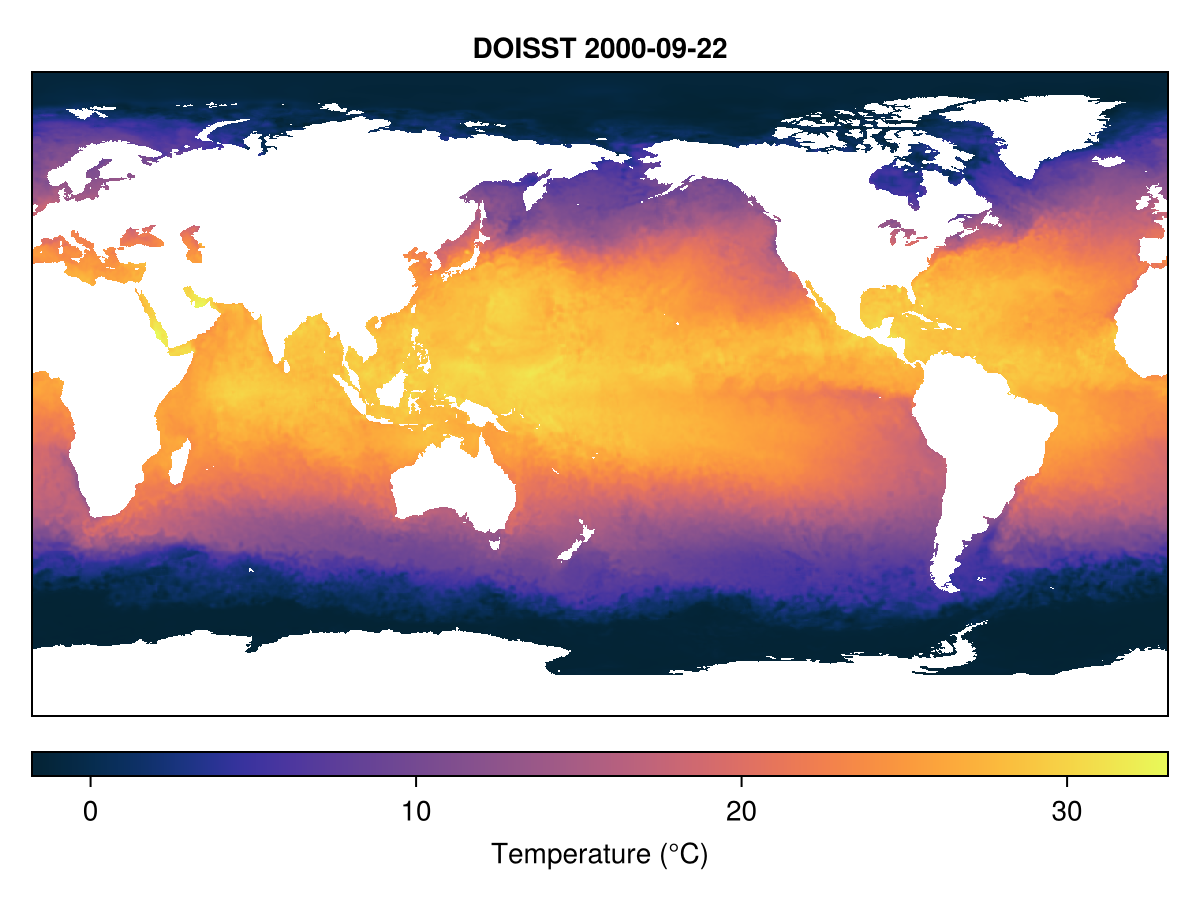}
\caption{Optimum interpolation sea surface temperature data for the
September Equinox of the year $2000$.}
\label{fig:sst_example}
\end{figure}

Figure~\ref{fig:sst_example} shows an example of the sea surface temperature
mesh data considered, this sample is for the September Equinox of the year $%
2000$.

Considering the importance of the oceans in climate dynamics, we hope this
data set allows us to extract important information about the evolution of
the climate over the period provided. For pedagogical analysis, we gather
the spatial data provided by the dataset into daily statistical time series.
Let us consider the quantity $T_{i,k}^{(j)}$, which denotes the temperature
at the $j$-th location (with $j=1,\dots ,N_{\text{place}}$) on day $%
k=1,\dots ,\Delta $ (with $\Delta \leq 365$) of year $t=1,\dots ,N_{\text{%
year}}$. The average temperature at location $j$ during year $t$ is then
defined as:

\begin{equation*}
\xi_{tj} = \left\langle T_t^{(j)} \right\rangle = \frac{1}{\Delta}
\sum_{k=1}^{\Delta} T_{t,k}^{(j)}
\end{equation*}

We denote by $P(\xi_{tj})$ the spatial distribution of these yearly averages
for year $t$. To standardize these values, we define the normalized variable:

\begin{equation*}
z_{tj}=\frac{\xi _{tj}-\overline{\xi _{t}}}{\sigma _{t}}
\end{equation*}%
where the spatial mean and standard deviation for year $t$ are given by:

\begin{equation*}
\overline{\xi_t} = \frac{1}{N_{\text{place}}} \sum_{j=1}^{N_{\text{place}}}
\xi_{tj}
\end{equation*}

\begin{equation*}
\sigma _{t}^{2}=\frac{1}{N_{\text{place}}-1}\sum_{j=1}^{N_{\text{place}%
}}\left( \xi _{tj}-\overline{\xi _{t}}\right) ^{2}\approx \overline{\xi
_{t}^{2}}-\overline{\xi _{t}}^{2}
\end{equation*}

In addition to this, we examine the temporal evolution of the mean $%
\overline{\xi _{t}}$ and variance $\sigma _{t}^{2}$, as well as the skewness
and kurtosis of the original (non-standardized) distributions $P(\xi _{tj})$%
, defined respectively as \cite{Trivedi2002}:

\begin{equation*}
S(t) = \overline{z_{tj}^3} = \frac{1}{N_{\text{place}}} \sum_{j=1}^{N_{\text{%
place}}} \left( \frac{\xi_{tj} - \overline{\xi_t}}{\sigma_t} \right)^3
\end{equation*}

\begin{equation*}
K(t)=\overline{z_{tj}^{4}}=\frac{1}{N_{\text{place}}}\sum_{j=1}^{N_{\text{%
place}}}\left( \frac{\xi _{tj}-\overline{\xi _{t}}}{\sigma _{t}}\right) ^{4}
\end{equation*}

We then considered it interesting to analyze the distribution $P(\xi _{tj})$,
which reflects the standardized spatial distribution of average temperatures
for the year $t$. A key question is whether these distributions
$P(\xi_{1j})\allowbreak,
P(\xi_{2j})\allowbreak,\dots\allowbreak,
P(\xi_{N_{\text{year}}j})$
as represented by their empirical histograms, exhibit universality across different years.
\ 

\begin{figure}[tbp]
\centering\includegraphics[width=0.9\linewidth]{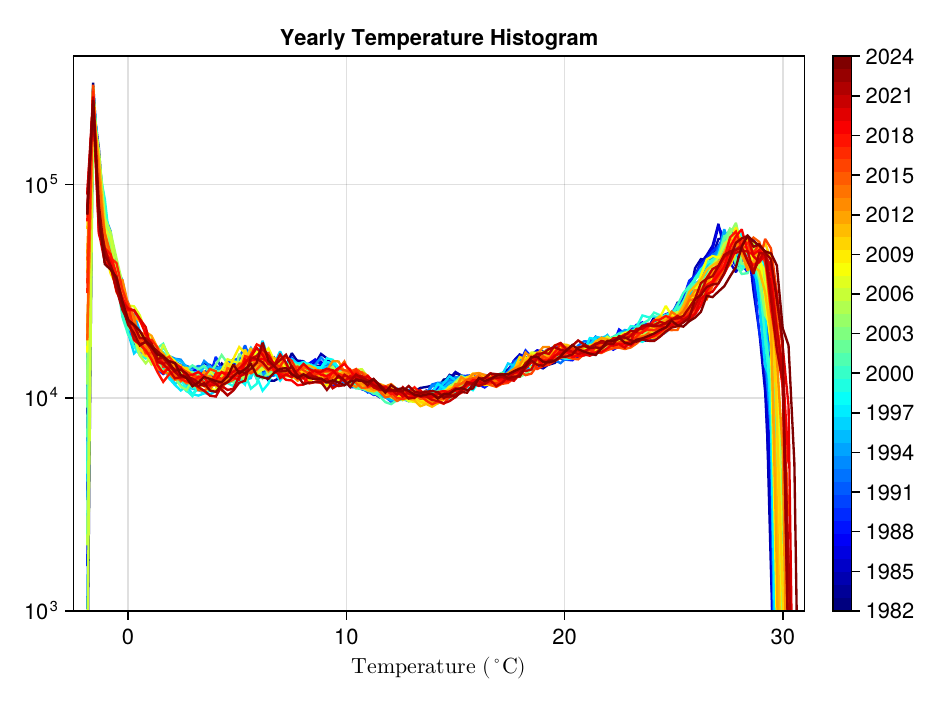}
\caption{Spatial histogram of yearly average temperature.}
\label{fig:global_hist_new}
\end{figure}

\begin{figure}[tbp]
\centering\includegraphics[width=0.9%
\linewidth]{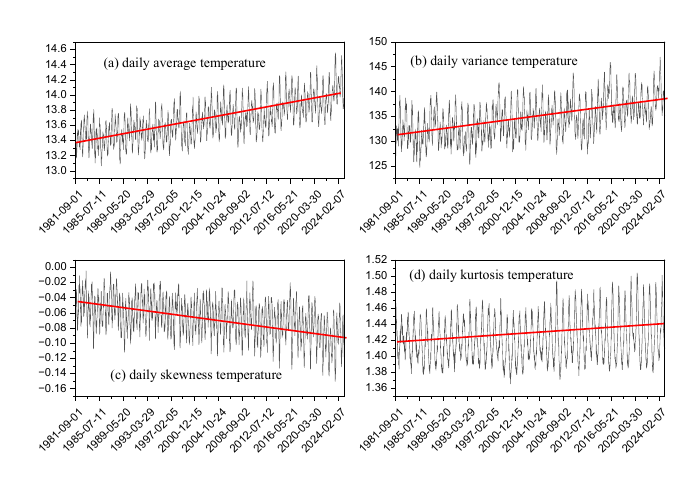}
\caption{Spatial statistics of the sea surface temperature.}
\label{fig:global_statistics}
\end{figure}

Figure~\ref{fig:global_hist_new} shows the spatial histogram of temperatures
averaged yearly. Although the distribution maintains its overall shape
(universal shape), the more recent distributions (in red) exhibit an
intensification of the aforementioned characteristics: a shift towards
higher temperatures and an increasing disparity between warm and cold
regions. However, such conclusions must be reinforced by the statistics defined above. These higher-order moments allow us to characterize the shape of
the distributions. These measures $\overline{\xi _{t}}$, $\sigma _{t}^{2}$, $%
S(t)$, and $K(t)$ are shown in plots (a), (b), (c), and (d), respectively, of Figure~\ref{fig:global_statistics}.

We performed linear fits for each quantity, considering that the first is the 0-th day and the last one is 15850-th. The slopes for average, variance,
skewness, and kurtosis are respectively $(3.92\pm 0.03)\times 10^{-5}$, $%
(4.30\pm 0.06)\times 10^{-4}$, $-(2.43\pm 0.04)\times 10^{-6}$, and $%
(1.09\pm 0.05)\times 10^{-6}$. These spatial temperature statistics reveal
clear patterns of anomalous behavior. Specifically, the steady rise in
average temperatures confirms a global warming trend, while the increase in
spatial variance highlights a growing thermal disparity between the warm and
cold regions. Furthermore, the negative skewness (third central moment of
temperature) indicates a distribution skewed toward higher temperatures.
Finally, the kurtosis (fourth central moment), a measure of the
distribution's tailedness, suggests that extreme temperature events are
becoming increasingly frequent.

After the presentation of the dataset and this initial exploration, we can
now show how to extract the seasonality of our time series that is
necessary to perform the main study of this paper, the study of the spectra of eigenvalues of correlation matrices built with such series.

\subsection{Seasonality}

\label{Subsec:Seasonality}

The time series under consideration are derived from climate data and
therefore exhibit seasonality, which is a recurring annual pattern in their
behavior. A proper analysis of the correlations between these time series
across different years requires the removal of this seasonal component. This
is achieved for each series by calculating its discrete Fourier transform
(DFT) and filtering out the annual frequency, here we considered $f =
1/365.24$, and its corresponding harmonics from the spectrum. The
deseasonalized time series is then obtained by applying the inverse DFT to
the filtered spectrum. This process is illustrated in Figure \ref%
{fig:dft_seasonality}.

\begin{figure}[]
\centering
\begin{tabular}{cc}
\includegraphics[width=0.5\textwidth]{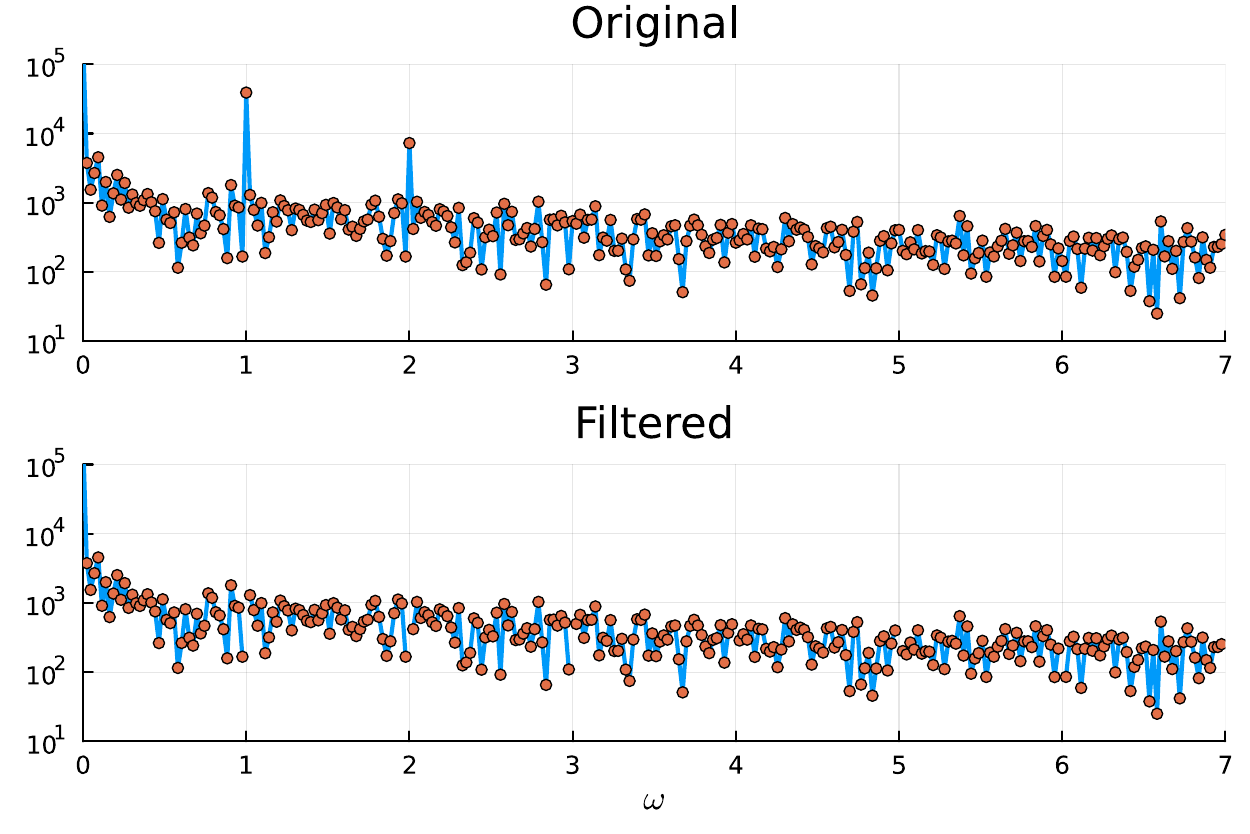} & %
\includegraphics[width=0.5\textwidth]{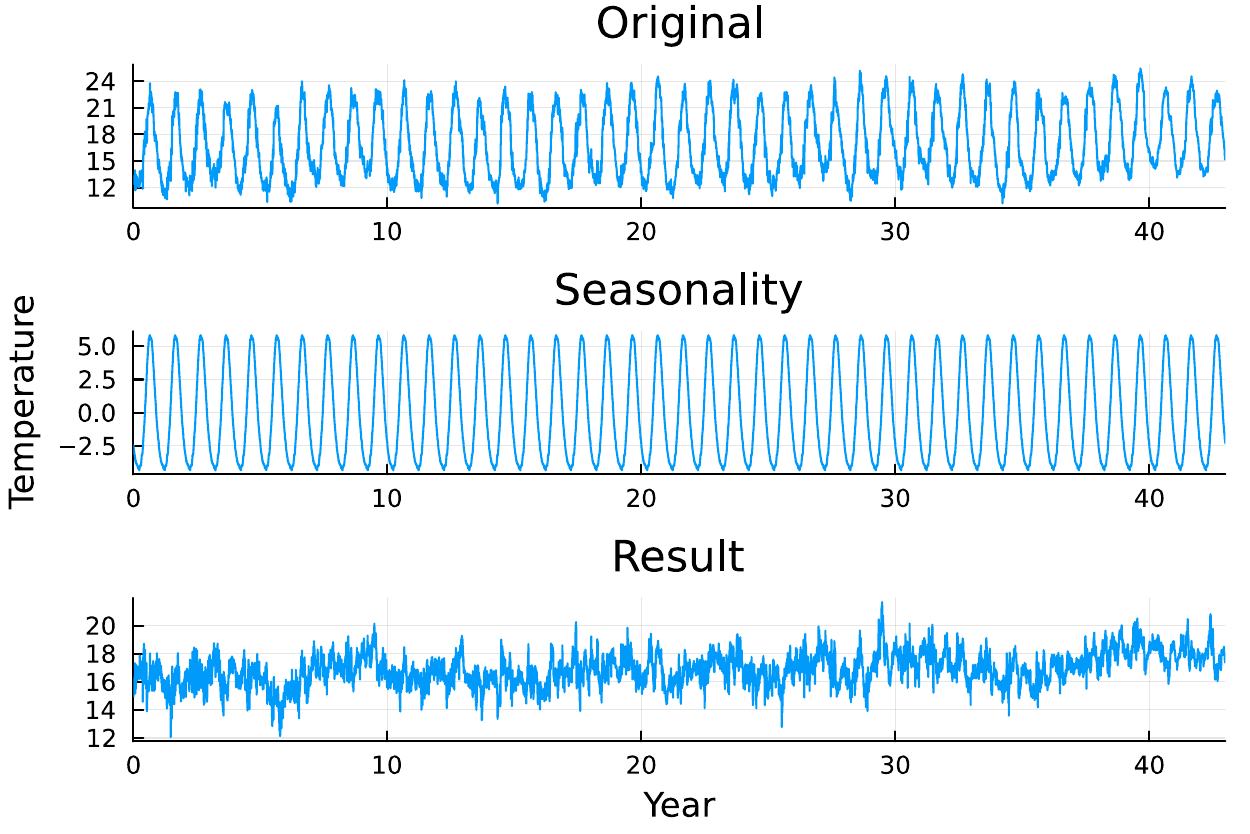} \\ \hline
&  \\ 
\includegraphics[width=0.5\textwidth]{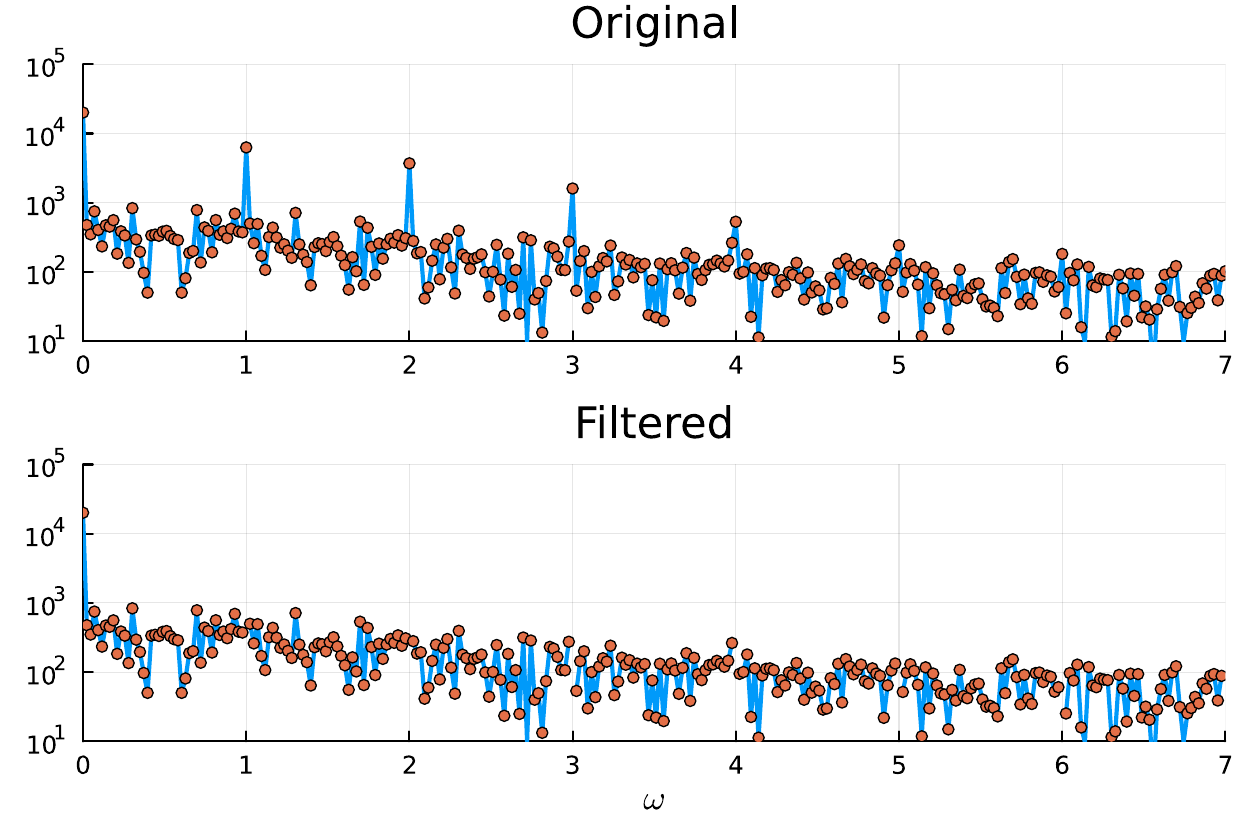} & %
\includegraphics[width=0.5\textwidth]{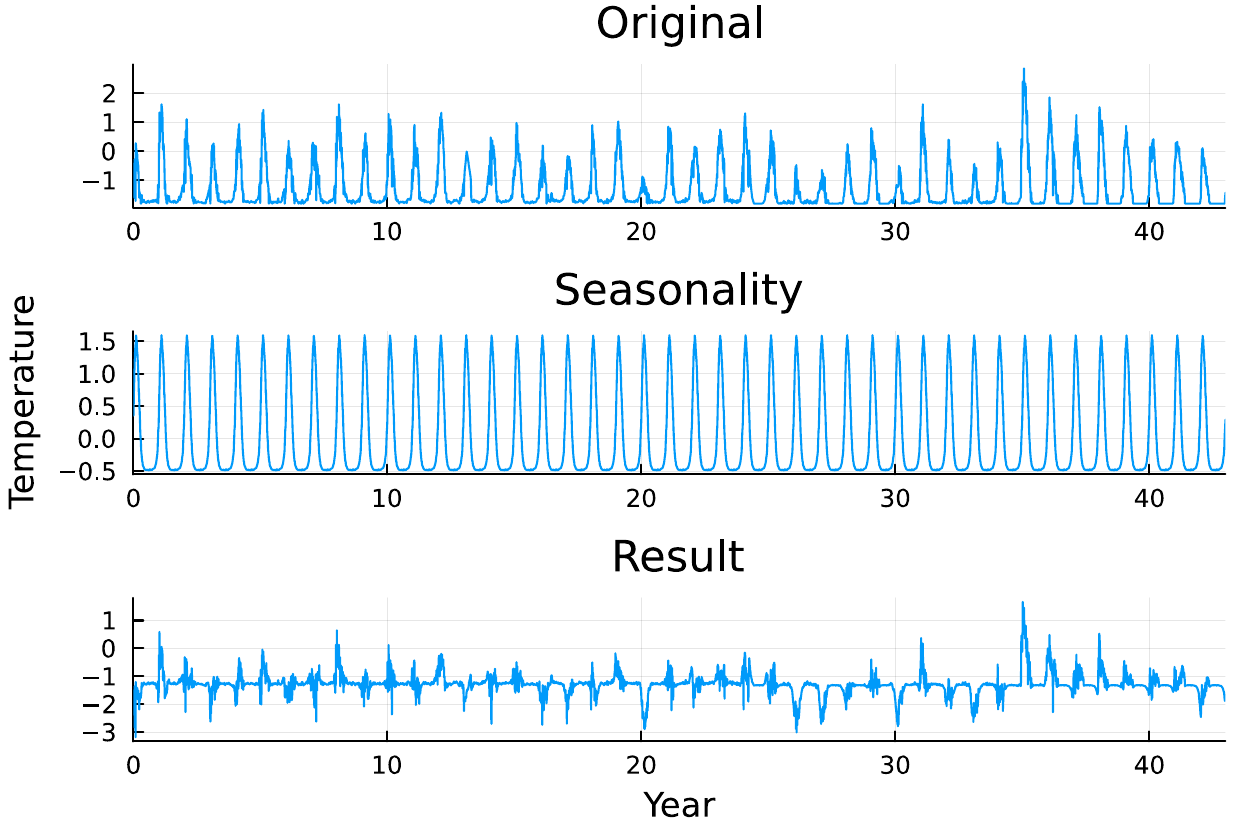} \\ 
& 
\end{tabular}%
\caption{ Illustration of the seasonality removal for daily surface
temperature time series for two different points on the ocean. The bottom
time series corresponds to a higher latitude than the top one. On the left,
we show the Fourier spectrum of the time series before and after the filter
is applied. We can clearly see local maxima on the yearly periodic component
and its harmonics on the original spectrum. On the right, we illustrate the
original time series, the removed seasonality component, and the resulting
time series. }
\label{fig:dft_seasonality}
\end{figure}

With the seasonality component removed, these time series are now ready for
use in our spectral analysis method to extract relevant information. In the
following section, we will describe the important points about correlation
matrices in the context of random matrix theory.

\section{Correlation random matrices}

\label{Sec:random_matrices_methodology}

The use of correlation random matrices requires the initial construction of
rectangular matrices with time series of certain quantities. We consider a
matrix $M$ whose $N_{\text{samples}}$ columns are composed of different time
series (in our case, temperature time series) samples each of length $N_{%
\text{steps}}$, where the matrix element $m_{ij}$ denotes the value of the $%
j $-th time series sample at the $i$-th time step. To analyze the spectral
properties, an interesting alternative is to consider the correlation matrix 
$G=\frac{1}{N_{\text{steps}}}M^{T}M$, which is a square matrix of size $N_{%
\text{samples}}\ \times N_{\text{samples}}$ with elements 
\begin{equation*}
g_{ij}=\frac{1}{N_{\text{steps}}}\sum_{k=1}^{N_{\text{steps}}}m_{ki}m_{kj}.
\end{equation*}%
It is crucial to note that $N_{\text{steps}}\ \geq N_{\text{samples}}$ in
all the following investigations.

It is advantageous to transform the components of the matrix $M$ into the matrix 
$M^{\ast }$, whose elements are standardized as follows: 
\begin{equation*}
m_{tj}^{\ast }=\frac{m_{tj}-\left\langle m_{j}\right\rangle _{t}}{\sqrt{%
\left\langle m_{j}^{2}\right\rangle _{t}-\left\langle m_{j}\right\rangle
_{t}^{2}}},
\end{equation*}%
where $\left\langle m_{j}^{k}\right\rangle _{t}=\frac{1}{N_{\text{steps}}}%
\sum_{i=1}^{N_{\text{steps}}}m_{ij}^{k}$. This transformation facilitates
subsequent analysis and calculations.

Therefore: 
\begin{equation}
g_{ij}^{\ast }=\frac{\left\langle m_{i}m_{j}\right\rangle -\left\langle
m_{i}\right\rangle \left\langle m_{j}\right\rangle }{\sigma _{i}\sigma _{j}},
\label{Eq:Correlation}
\end{equation}%
where $\left\langle m_{i}m_{j}\right\rangle _{t}=\frac{1}{N_{\text{steps}}}%
\sum_{t=1}^{N_{\text{steps}}}m_{ti}m_{tj}$ and $\sigma _{i}=\sqrt{%
\left\langle m_{i}^{2}\right\rangle -\left\langle m_{i}\right\rangle ^{2}}$.

In the general case, when the variables $m_{ij}^{\ast }$ are uncorrelated
random variables, the\ Wishart-like joint distribution of eigenvalues of
matrix $G^{\ast }=\frac{1}{N_{\text{steps}}}M^{\ast T}M^{\ast }$ is given by
a Boltzmann weight: 
\begin{equation}
\begin{array}{lll}
P_{W}(\lambda _{1},\cdots ,\lambda _{N_{\text{samples}}}) & = & C_{N_{\text{%
samples}}}\exp \left( -\frac{N_{\text{steps}}}{2}\sum_{i=1}^{N_{\text{samples%
}}}\lambda _{i}\right. \\ 
&  &  \\ 
&  & +\left. \frac{(N_{\text{steps}}\ -N_{\text{samples}}\ -1)}{2}%
\sum_{i=1}^{N_{\text{samples}}}\ln \lambda _{i}+\sum_{i<j}\ln \left\vert
\lambda _{i}-\lambda _{j}\right\vert \right) \\ 
&  &  \\ 
& = & C_{N_{\text{steps,}}N_{\text{samples}}}\exp (-\beta \mathcal{H}%
(\lambda _{1},...,\lambda _{N_{\text{samples}}}))%
\end{array}
\label{Eq:joint} \\
\end{equation}%
with Hamiltonian: 
\begin{equation*}
\mathcal{H}(\lambda _{1},...,\lambda _{N_{\text{samples}}})=\sum_{i=1}^{N_{%
\text{samples}}}V(\lambda _{i})-\sum_{i<j}\ln \left\vert \lambda
_{i}-\lambda _{j}\right\vert \text{,}
\end{equation*}%
where: $V(\lambda )=\frac{N_{\text{steps}}}{2}\lambda -\frac{(N_{\text{steps}%
}\ -N_{\text{samples}}\ -1)}{2}\ln \lambda $ and $\beta =1.$ Here $C_{N_{%
\text{steps,}}N_{\text{samples}}}=Z^{-1}$ is the normalization, where: 
\begin{equation*}
Z=\int_{-\infty }^{\infty }\cdots \int_{-\infty }^{\infty }\exp (-\beta 
\mathcal{H}(\lambda _{1},...,\lambda _{N_{\text{samples}}}))d\lambda
_{1}\cdots d\lambda _{N_{\text{samples}}}
\end{equation*}

If we integrate the $P_{W}(\lambda _{1},\cdots ,\lambda _{N_{\text{samples}%
}})$ in all eigenvalues except by one

\begin{equation}
\sigma (\lambda )=\int_{-\infty }^{\infty }\cdots \int_{-\infty }^{\infty
}d\lambda _{1}\cdots d\lambda _{N_{\text{samples}}-1}\ P_{W}(\lambda
_{1},\cdots ,\lambda _{N_{\text{samples}}\ -1},\lambda ),
\label{eq:integration}
\end{equation}%
we obtained the well-known MP distribution~\cite%
{marchenko1967distribution}: 
\begin{equation}
\sigma (\lambda )=%
\begin{cases}
\frac{N_{\text{steps}}}{2\pi N_{\text{samples}}}\frac{\sqrt{(\lambda
-\lambda _{-})(\lambda _{+}-\lambda )}}{\lambda }, & \text{if}\ \lambda
_{-}\leq \lambda \leq \lambda _{+} \\ 
0, & \text{otherwise},%
\end{cases}
\label{eq:marchenko-pastur}
\end{equation}%
where $\lambda _{\pm }=1+\frac{N_{\text{samples}}}{N_{\text{steps}}}\pm 2%
\sqrt{\frac{N_{\text{samples}}}{N_{\text{steps}}}}$, corresponding to a
stable law for the density of eigenvalues. Deviations from this law are then
expected for the correlated time series since the joint distribution of
eigenvalues does not follow Eq. \ref{Eq:joint} and therefore the density of
eigenvalues must escape from the stable law described by the MP law
(Eq. \ref{eq:marchenko-pastur}). For financial time series \cite%
{stanleyQuantifyingFluctuationsEconomic2000,bouchaudFinancialRMT2009}
outliers to $\lambda _{+}=1+\frac{N_{\text{samples}}}{N_{\text{steps}}}+2%
\sqrt{\frac{N_{\text{samples}}}{N_{\text{steps}}}}$ are verified by
characterizing genuine correlations for stock markets, however we verified
very distinct behavior for the density of eigenvalues for the climate time
series as we will observe.

\section{Spectral analysis}

\label{Sec:Spectral_Analysis}

To perform the spectral analysis, we uniformly sample $N_{\text{place}}\
=10^{5}$ locations from the ocean mesh and extract the corresponding daily
sea surface temperature time series, as illustrated in Figure \ref%
{fig:noaa_ts_extraction}. The resulting time series are then grouped by
year. For each year, the data are partitioned into $N_{\text{runs}}\ =1000$
groups, each containing $N_{\text{samples}}\ =100$ time series. Each time
series consists of $N_{\text{steps}}=365$ daily measurements (366 for leap
years).

\begin{figure}[tbp]
\centering
\includegraphics[width=0.9\linewidth]{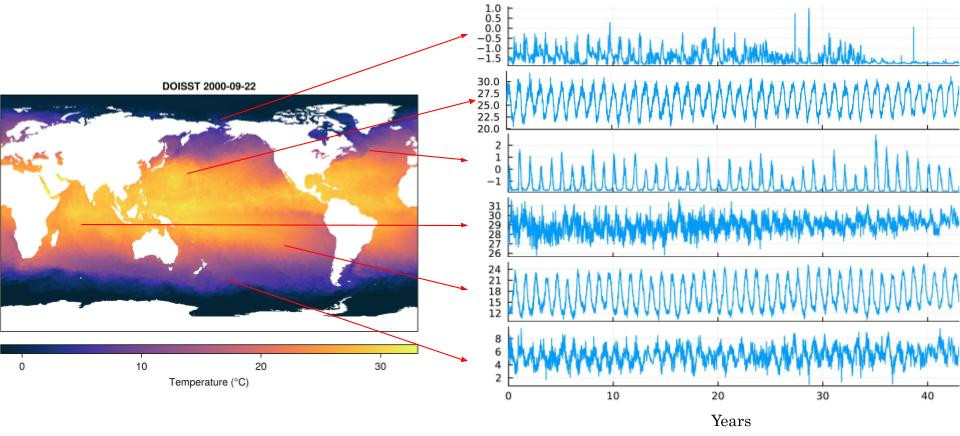}
\caption{Daily surface temperature time series are sampled from uniformly
distributed random points over the ocean surface. Note that this figure is
intended solely to illustrate the sampling procedure and does not correspond
to the actual coordinates used in the analysis. }
\label{fig:noaa_ts_extraction}
\end{figure}

\begin{figure}[tbp]
\centering\includegraphics[width=0.9\linewidth]{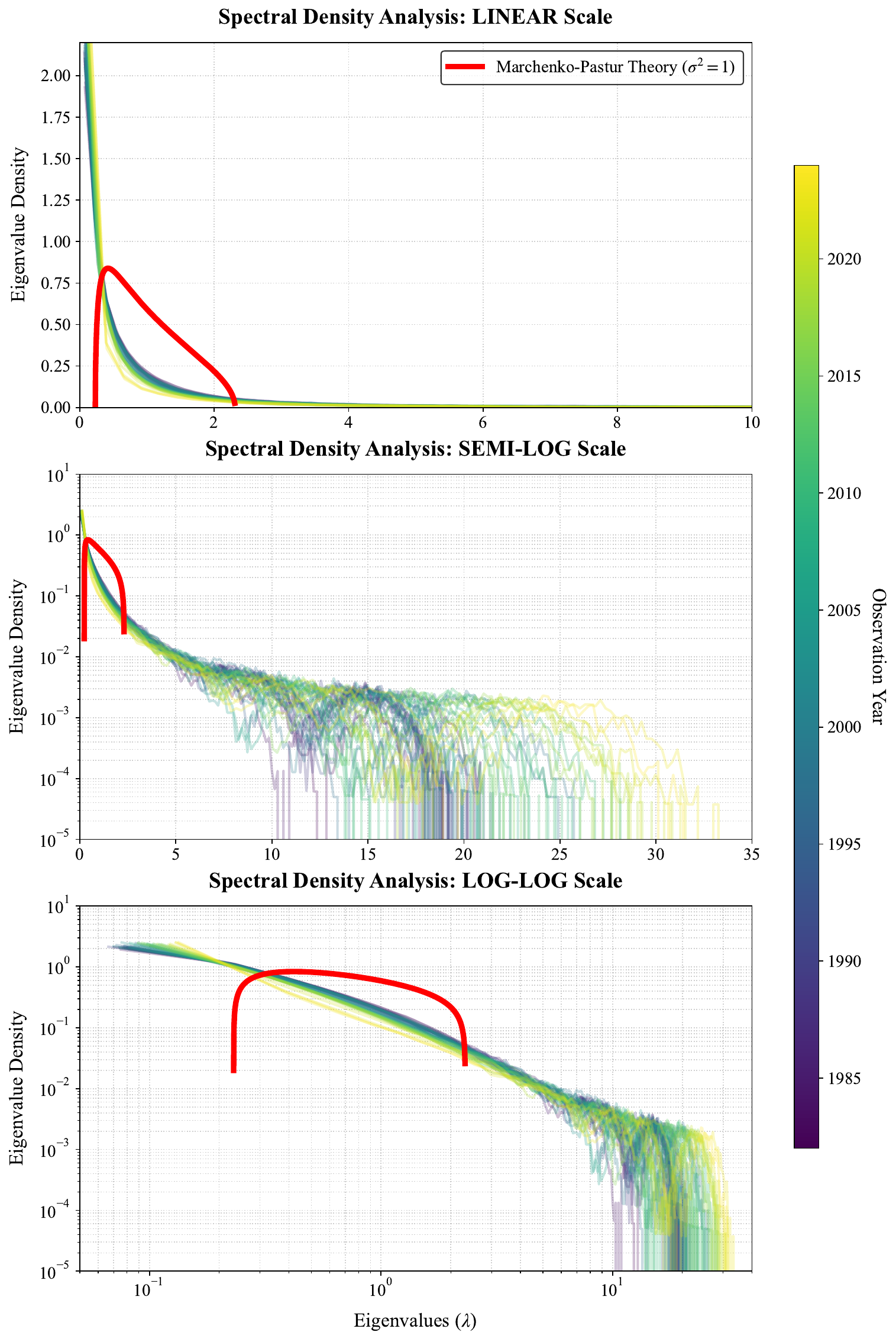}
\caption{Yearly histograms of the eigenvalues of the correlation matrices
constructed from the sea surface temperature time series. The
MP distribution is shown in red for reference. Linear,
semi-logarithmic, and log-log scales are used to highlight different regions
of the eigenvalue density. Data from different years are displayed in
different colors.}
\label{fig:sst_global_eigvals_hist}
\end{figure}

Each group is used to construct a data matrix from which a correlation
matrix is computed. This procedure generates an ensemble of $N_{\text{runs}}$
correlation matrices for each year, enabling the investigation of the
temporal evolution of their spectral properties. After removing the seasonal
component, as described in Subsection \ref{Subsec:Seasonality}, we applied
this ensemble-based approach to estimate the eigenvalue density and other
spectral statistics. Figure \ref{fig:sst_global_eigvals_hist} shows the
eigenvalue density for selected years and compares the results with the
MP distribution.

We can clearly observe that the spectrum differs markedly from the
MP distribution, with a substantial number of
eigenvalues lying well above the maximum predicted by this theoretical
bound, indicating the presence of strong correlations among the climate time
series. Moreover, the eigenvalue density exhibits pronounced tails whose
structure appears to depend on the year under consideration, suggesting that
the spectral properties are sensitive to long-term trends associated with
global warming. As expected, the eigenvalue distribution deviates
significantly from the MP prediction, since nontrivial correlations persist
even after the removal of seasonal components.

On both semi-logarithmic and log-log scales, a significant number of
eigenvalues are found beyond the upper edge predicted by the bulk
distribution. Rather than appearing as isolated fluctuations, these large
eigenvalues form a well-defined cluster whose characteristic position shifts
with the year under consideration. This behavior suggests the presence of
strong collective correlations that are not captured by the standard
random-matrix description of the bulk spectrum.

\begin{figure}[tbp]
\centering\includegraphics[width=0.9\linewidth]{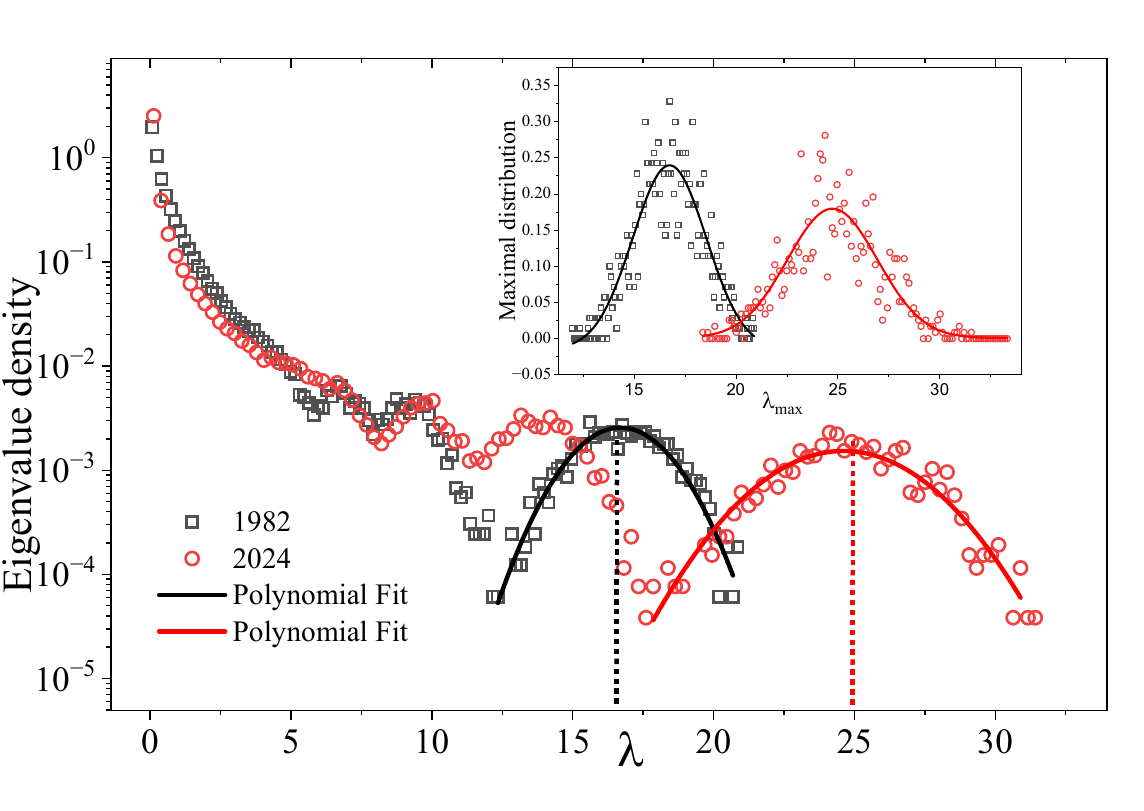}
\caption{Comparison of the maximum-eigenvalue distributions for 1982 and
2024. The tails are well described by Gaussian distributions, indicating
that the fluctuations of the largest eigenvalue are approximately Gaussian
in both cases. However, the distributions are centered at markedly different
mean values, reflecting the temporal evolution of the system. The inset
displays the complete distributions of the maximum eigenvalue on a linear
scale, confirming the same qualitative behavior observed in the tail
analysis and highlighting the shift of the distribution toward larger
eigenvalues in 2024.}
\label{Fig:Maximum}
\end{figure}

To investigate this feature more thoroughly, we analyze the statistics of
the largest eigenvalue. In particular, we focus on the years 1982 and 2024,
which provide representative examples of the evolution observed throughout
the dataset. The corresponding distributions of the maximum eigenvalue are
shown in Fig. \ref{Fig:Maximum}. Remarkably, the tails are well described by
Gaussian functions, which appear as parabolic profiles in a semi-logarithmic
representation. The Gaussian character of these distributions contrasts with
the Tracy--Widom behavior expected for weakly correlated Wishart ensembles
and provides further evidence that strong correlations play a dominant role
in shaping the spectral properties of the system \cite{Johnstone2000}.

The inset displays the full distribution of the largest eigenvalue, rather
than only its tail, and reveals the same qualitative features observed in
the tail analysis. For clarity, the distribution is shown on a linear scale.
The observed Gaussian behavior, therefore, suggests that the system departs
from the usual universal regime.

\begin{figure}[tbp]
\centering
\includegraphics[width=0.9\linewidth]{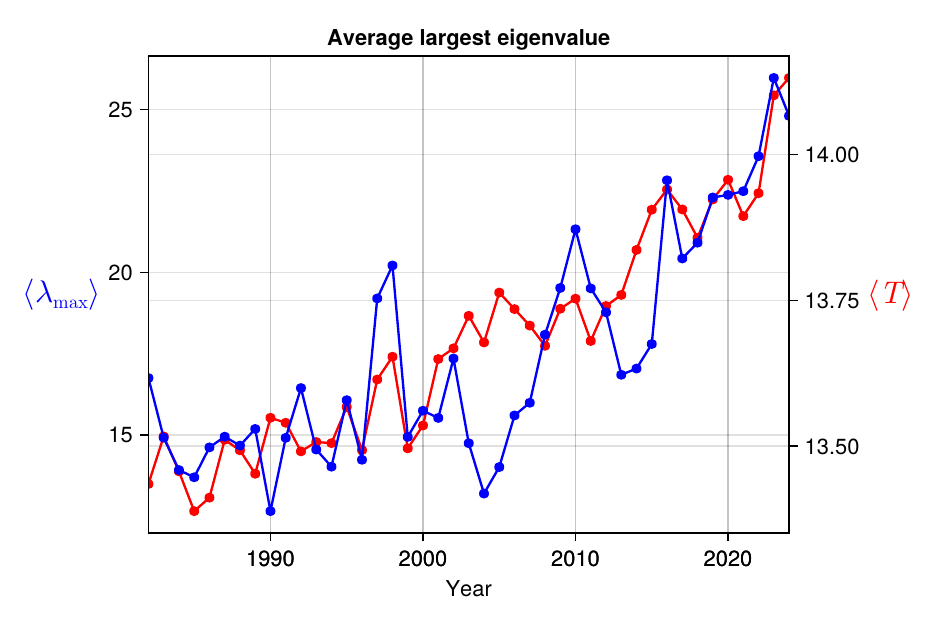}
\caption{Average value of the largest eigenvalue of the yearly correlation
matrices as a function of time, compared with the corresponding global mean
temperature. Both quantities exhibit a clear upward trend throughout the
analyzed period, highlighting a strong association between the evolution of
the dominant correlation mode and the increase in global temperature.}
\label{Fig:sst_eigval_max_avg}
\end{figure}

Nevertheless, a clear increasing trend is observed when considering the
maximum eigenvalue of each yearly correlation matrix, as illustrated in
Figure \ref{Fig:sst_eigval_max_avg}. This trend curiously parallels the
global average temperature. We speculate that this phenomenon is a
consequence of global warming, as the largest eigenvalue is linked to the
strongest collective mode within the ensemble of time series, representing
the warming trend.

To better explore the effects of the warming global, we go beyond
by looking at the yearly evolution of the average $n$-th eigenvalue (Fig. \ref%
{fig:sst_eigvals_ts}) which can be compared against the support of the MP
distribution for scale. For all years considered, the spread of the spectra
exceeds the range predicted by the MP distribution.

\begin{figure}[tbp]
\centering\includegraphics[width=0.9\linewidth]{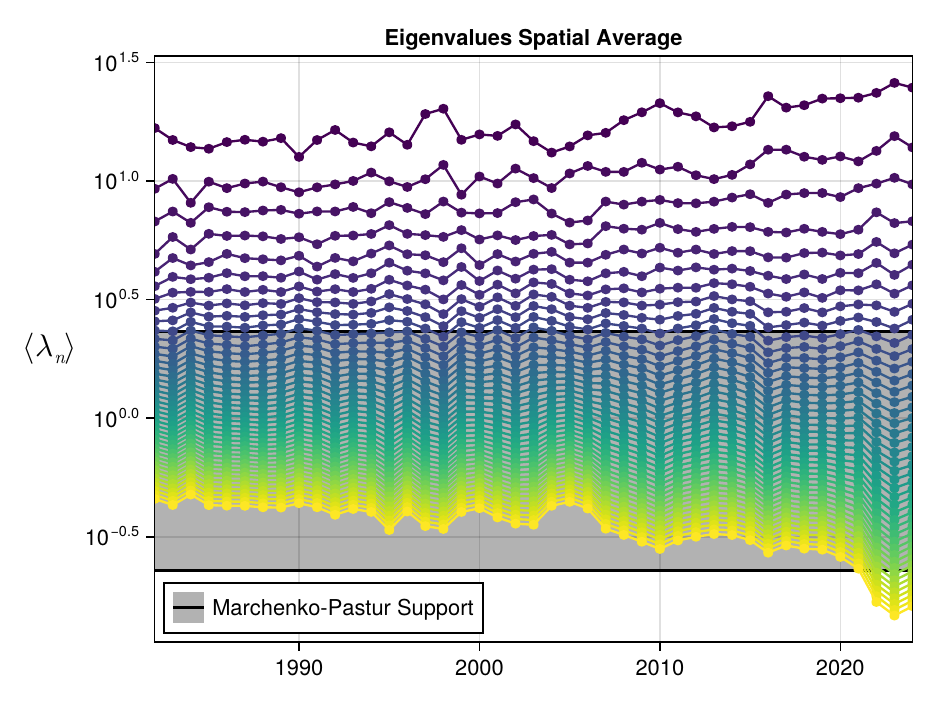}
\caption{Average value of the $n$-th eigenvalue of the yearly correlation
matrices, ordered by decreasing magnitude. The shaded gray region indicates
the support of the MP distribution and serves as a reference
for the range expected in the absence of significant correlations.}
\label{fig:sst_eigvals_ts}
\end{figure}

These findings extend the application of correlation random-matrix methods
beyond their traditional use in spin models and other theoretical complex
systems \cite%
{daSilvaRandomMatricesTheory2023,daSilva2023meanField,nosso_article_entropy_2024,daSilva2024PottsModel,daSilva2025Identifying,daSilva2025RevisitingContact,daSilva2023Dynamics}%
. They further demonstrate the ability of spectral techniques to
characterize the dynamics of real-world complex systems, as previously
established in financial markets \cite%
{stanleyQuantifyingFluctuationsEconomic2000,bouchaudFinancialRMT2009}, and
now, in the present work, in the context of climate change.

By successfully identifying stylized facts associated with climate
variability and global warming using relatively small correlation matrices,
we further highlight the versatility and robustness of this spectral
framework. The results presented here suggest several promising directions
for future research. In particular, a more localized analysis focusing on
specific oceanic regions and their interactions with nearby countries or
territories may provide valuable insights into regional climate dynamics.
Such studies could help clarify possible connections between evolving
ocean-temperature correlations and the increasing frequency and intensity of
extreme climate events, including those that have recently affected
southeastern and especially southern Brazil. These possibilities illustrate
the broader potential of random-matrix-based approaches as tools for
investigating climate-related phenomena across multiple spatial and temporal
scales.

\section{Summary and Conclusions}

\label{Sec:Conclusions}

We investigated correlation random matrices constructed from ocean
temperature data to characterize signatures of climate change
through their spectral properties. Our results reveal a strong relationship
between global warming and the behavior of the largest eigenvalue. In
particular, the average maximum eigenvalue exhibits a clear increasing trend
over time, closely following the rise in global ocean temperatures.

From the perspective of random matrix theory, we find that the empirical
eigenvalue density deviates significantly from the prediction of the
MP law, indicating the presence of strong correlations in the
underlying climate system. Moreover, the distribution of the largest
eigenvalue is found to be approximately Gaussian rather than Tracy--Widom,
suggesting that the system lies outside the universal regime typically
associated with weakly correlated Wishart ensembles. These findings provide
additional evidence that climate-driven correlations become increasingly
relevant as the system evolves.

The spectral results are further supported by the behavior of conventional
statistical descriptors of the temperature field, including the variance,
skewness, and kurtosis. Together, these measures consistently indicate
substantial changes in the statistical structure of ocean temperatures over
the period analyzed.

Overall, our study demonstrates that important signatures of climate change
can be detected through both the eigenvalue spectrum and the statistics of
extremal eigenvalues of correlation matrices. These preliminary but
encouraging results highlight the potential of random matrix methods as
complementary tools for climate data analysis and motivate further
investigations into their applicability to a broader range of climatic
variables and datasets.

\textbf{Funding:} RDS acknowledges financial support from the Brazilian agency CNPq (Conselho Nacional de Desenvolvimento Científico e Tecnológico) under Grants 304575/2022-4 and 309560/2025-0. RDS, EVS, and SG also thank FAPERGS (Fundação de Amparo à Pesquisa do Estado do Rio Grande do Sul) for support under Grant 25/2551-0002529-0.

\bibliographystyle{elsarticle-num}
\bibliography{references}

\end{document}